\documentstyle[aps,version2,pre]{revtex}    
\begin{document}

\draft

\begin{title}
Instabilities of the Hubbard chain in a magnetic field
\end{title}

\author{J. M. P. Carmelo$^{1,2}$, F. Guinea$^{2}$, and 
P. D. Sacramento$^{3,*}$}
\begin{instit}
$^{1}$ Department of Physics, University of \'Evora,
Apartado 94, P-7001 \'Evora Codex, Portugal and\\
Centro de F\'{\i}sica das Interac\c{c}\~oes Fundamentais,  
I.S.T., P-1096 Lisboa Codex, Portugal
\end{instit}
\begin{instit}
$^{2}$ Instituto de Ciencia de Materiales, C.S.I.C.,
Cantoblanco, E-28949 Madrid, Spain
\end{instit}
\begin{instit}
$^{3}$ Theoretical Physics, University of Oxford, 1 Keble Road, Oxford,
OX1 3NP, UK
\end{instit}
\receipt{September 1996}

\begin{abstract}
We find and characterize the instabilities of the repulsive Hubbard
chain in a magnetic field by studing all response functions
at low frequency $\omega $ and arbitrary momentum. The instabilities
occur at momenta which are simple combinations of the ($U=0$)
$\sigma =\uparrow ,\downarrow$ Fermi points, $\pm k_{F\sigma}$.
For finite values of the on-site repulsion $U$ the instabilities 
occur for single $\sigma $ electron adding or removing at momenta 
$\pm k_{F\sigma}$, for transverse spin-density wave (SDW) at momenta 
$\pm 2k_F$ (where $2k_F=k_{F\uparrow}+k_{F\downarrow}$), and 
for charge-density wave (CDW) and SDW at momenta $\pm 2k_{F\uparrow}$
and $\pm 2k_{F\downarrow}$. While at zero magnetic field removing or 
adding single electrons is dominant, the presence of that 
field brings about a dominance for the transverse $\pm 2k_F$ SDW over 
all the remaining instabilities for a large domain of $U$ and 
density $n$ values. We go beyond conformal-field theory and 
study divergences which occur at finite frequency in the one-electron 
Green function at half filling and in the transverse-spin response 
function in the fully-polarized ferromagnetic phase.
\end{abstract}
\renewcommand{\baselinestretch}{1.656}   


\section{INTRODUCTION}

One-dimensional (1D) electronic quantum problems have peculiar 
instabilities which follow from their restrictive geometry.
For instance, it is well known that the low-frequency charge and 
spin response functions of a gas of electrons in a 1D lattice 
show logarithmic divergences at twice the Fermi wavevector. This is 
the so called Peierls instability which in real quasi-one-dimensional
materials \cite{Pouget,Kago} occurs through coupling of the electrons 
to suitable phonon modes giving rise to lattice distortions and to 
phenomena like charge-density waves (CDW's) or spin-density waves 
(SDW's).

These instabilities also occur in the presence of a magnetic
field and, in real materials there are electron - electron
interactions which at low dimensions have a strong impact
on the physical properties. In this paper we study the effects
of the interplay between a magnetic field and many-electron 
interactions on the 1D quantum-problem instabilities. We consider 
an integrable many-electron problem \cite{Yang,Korepinrev}, 
the Hubbard chain \cite{Lieb} in a magnetic field 
\cite{Frahm,Frahm91,Ogata}. One of our goals is to use 
elsewhere the present results on the 1D instabilities as a
starting point for the study of the corresponding instabilities
of a system of weakly coupled chains. Therefore, we include in
our present 1D study the one-particle Green function which
involves excitations that change the electron numbers. In 
the coupled-chain system this 1D function will provide
important information on the relevance of interchain electron 
hopping. In Sec. VI we will return to the problem of
the Hubbard coupled-chain system in a magnetic field.

At zero onsite interaction, $U=0$, the Hubbard chain
reduces to a 1D tight-binding electron model with first-neighbor 
transfer integral $t$. In this case it is easy to evaluate the 
low-frequency response functions $\hbox{Re}\chi^{\vartheta } 
(k,\omega)$ (with $\vartheta $ being charge $\rho$, spin projection 
$\sigma_z$, and so on) and to characterize and detect the 
corresponding divergences. At zero magnetic field the charge 
function $\hbox{Re}\chi^{\rho} (\pm 2k_F,\omega)$, the spin 
function $\hbox{Re}\chi^{\sigma_z} (\pm 2k_F,\omega)$, 
the spin-transverse function $\hbox{Re}\chi^{\sigma_{\perp}} 
(\pm 2k_F,\omega)$, the singlet superconductivity functions 
$\hbox{Re}\chi^{ss} (0,\omega)$ and $\hbox{Re}\chi^{ss} 
(\pm 2k_F,\omega)$, and the triplet $\sigma $ superconductivity 
function $\hbox{Re}\chi^{ts\sigma} (0,\omega)$ all diverge as $- ln 
(\omega)$. If we apply a magnetic field we find that the
$H=0$ instabilities are limiting cases of the finite-field
situation -- for $0<H<{2t\over \mu_0}[\sin ({\pi\over 2}n)]^2$ 
(here $H_c={2t\over \mu_0}[\sin ({\pi\over 2}n)]^2$ is the $U=0$
critical field for onset of fully-polarized ferromagnetism
and $\mu_0$ is the Bohr magneton) the charge functions 
$\hbox{Re}\chi^{\rho} (\pm 2k_{F\uparrow},
\omega)$ and $\hbox{Re}\chi^{\rho} (\pm 2k_{F\downarrow},\omega)$, 
the spin functions $\hbox{Re}\chi^{\sigma_z} (\pm 2k_{F\downarrow},\omega)$ 
and $\hbox{Re}\chi^{\sigma_z} (\pm 2k_{F\uparrow},\omega)$, the 
spin-transverse function $\hbox{Re}\chi^{\sigma_{\perp}} 
(\pm 2k_F,\omega)$, the singlet superconductivity functions 
$\hbox{Re}\chi^{ss} (\pm [k_{F\uparrow}-k_{F\downarrow}],\omega)$ 
and $\hbox{Re}\chi^{ss} (\pm 2k_F,\omega)$, and the triplet $\sigma $ 
superconductivity function $\hbox{Re}\chi^{ts\sigma} (0,\omega)$ 
all diverge as $- ln (\omega)$. Also, both for $H=0$ and $H>0$ the 
single-electron Green function diverges at $\pm k_{F\sigma}$ as 
$1/\omega$.

On the other hand, the study of all divergences of the low-frequency 
response functions $\hbox{Re}\chi^{\vartheta } (k,\omega)$ is
for the corresponding many-electron quantum system a problem of
some complexity. The BA solution and conformal-field theory (CFT) 
can be combined to evaluate the asymptotics of the correlation 
functions of the Hubbard chain in a magnetic field
\cite{Frahm,Frahm91,Ogata}. However, the momentum-dependent 
expressions derived in Ref. \cite{Ogata} (for zero magnetic field 
see also Ref. \cite{Schulz}) and some of
the quantities studied in Ref. \cite{Frahm91} refer
to the equal-time situation and we need here the low-frequency
response functions. Moreover, Ref. \cite{Frahm91} evaluated
and presented expressions for the response functions which 
refer to particular relations between the values of the 
low-frequency $\omega $ and small-momentum $(k-k_0)$. Here $k_0$
are particular combinations of the $U=0$ electronic Fermi
momenta [see Eq. $(12)$ below]. We find that the low-frequency 
correlation functions $\hbox{Re}\chi^{\vartheta } (k,\omega)$ 
{\it can} only diverge at the particular values $k=k_0$. 
Unfortunately, Ref. \cite{Frahm91} has not solved the present 
problem of the instabilities because its expression $(5.7)$ 
corresponds to a very particular relation between $\omega $ and 
$(k-k_0)$ which {\it does not} provide the relevant and general 
$\omega $ dependence {\it at} $k=k_0$.

The problem was solved for the particular case of the charge
$\vartheta =\rho$ and spin $\vartheta =\sigma_z$ response
functions in Ref. \cite{Carmelo93}. However, that paper has 
not considered all correlation functions. Although the charge 
and spin excitations studied in Ref. \cite{Carmelo93} are relevant
and correspond to real instabilities, they are never
dominant, as we show in this paper. 

Using the same methods as in Refs. \cite{Carmelo93,Carmelo96b},
we find that all instabilities correspond at finite values  
of $U$ to power-law divergences (and in some singular points of
the parameter space to logarithmic divergences) in the correlation functions.
The corresponding exponents are non-classic combinations
of two-pseudoparticle phase shifts. These pseudoparticles
are the quantum objects introduced in Refs.
\cite{Carmelo90,Carmelo91,Carmelo92,Carmelo92b,Carmelo92c} which 
refer to the operator algebra \cite{Carmelo94,Carmelo94b,Carmelo96}
which diagonalizes the quantum problem. Their zero-momentum
forward-scattering interactions control the overlaps
between the Hamiltonian eigenstates and the excitations
associated with the correlation functions 
\cite{Carmelo93,Carmelo96b}. It is the exotic character of
such overlap which leads to the Luttinger-liquid
\cite{Haldane} non-classic exponents.

While at zero magnetic field removing or adding single $\sigma $ 
electrons at $k=\pm k_{F\sigma }$ is dominant (in the weakly 
coupled-chain problem this is associated with inter-chain hopping)
and, due to the spin $SU(2)$ symmetry, the transverse $\pm 2k_F$ 
SDW and the $\pm 2k_F$ SDW and CDW have the same $\omega$ dependence, the
presence of a magnetic field brings about a dominance for the 
transverse $\pm 2k_F$ SDW over the remaining instabilities
for a large domain of $U$ and $n$ values. There are no divergences 
in the superconductivity and other response functions. 

We study the effects of the half-filled metal -- insulator
transition and fully-polarized ferromagnetic transition
on the above correlation functions. Some low-energy power-law 
divergences are present in both sides of these transitions but 
there occur discontinuities in the corresponding non-classical 
critical exponents. Moreover, we go beyond CFT and study power-law 
divergences which occur for some correlation functions at finite energy
for the half-filling and fully-polarized ferromagnetic phases. 

The paper is organized as follows: the Hamiltonian and
the pseudoparticle operator basis are introduced in Sec. II. 
In Sec III we characterize the instabilities
of the quantum problem. The comparative study of these
instabilities is presented in Sec. IV. In Sec. V
we consider the half-filling and fully-polarized
ferromagnetic phases and the effects of the half-filling
metal - insulator transition and fully-polarized ferromagnetic
transitions on the quantum-problem instabilities.
The coupled-chain problem in the presence of a magnetic field
is discussed in Sec. VI.  Finally, Sec. VII presents the 
concluding remarks.

\section{THE HUBBARD CHAIN AND THE PSEUDOPARTICLE OPERATOR BASIS}

We consider the Hubbard chain \cite{Lieb,Frahm,Frahm91}
with a finite chemical potential $\mu$ and in the presence of 
a magnetic field $H$ \cite{Carmelo92b,Carmelo94,Carmelo94b}

\begin{eqnarray}
\hat{H} = -t\sum_{j,\sigma}\left[c_{j,\sigma}^{\dag }c_{j+1,\sigma}+c_
{j+1,\sigma}^{\dag }c_{j,\sigma}\right] +
U\sum_{j} [c_{j,\uparrow}^{\dag }c_{j,\uparrow} - 1/2]
[c_{j,\downarrow}^{\dag }c_{j,\downarrow} - 1/2]
+ 2\mu{\hat{\eta}}_z + 2\mu_0 H{\hat{S}}_z \, ,
\label{hamil}
\end{eqnarray}
where $c_{j,\sigma}^{\dag }$ and $c_{j,\sigma}$ are the creation and
annihilation operators, respectively, for electrons at the 
site $j$ with spin projection $\sigma=\uparrow, \downarrow$ and 
${\hat{\eta}}_z=-{1\over 2}[N_a-\sum_{\sigma}{\hat{N}}_{\sigma}]$
and ${\hat{S}}_z=-{1\over 2}\sum_{\sigma}\sigma{\hat{N}}_{\sigma}$.
Here ${\hat{N}}_{\sigma}$ is the $\sigma$-electron number operator.
We introduce the $U=0$ Fermi points $k_{F\sigma}=\pi n_{\sigma}$ (and 
$k_F=[k_{F\uparrow}+k_{F\downarrow}]/2=\pi n/2$) where 
$n_{\sigma}=N_{\sigma}/N_a$ and $n=N/N_a$, and  
$N_a$ is the number of lattice sites. The total-electron number
is $N=\sum_{\sigma}N_{\sigma}$ and the spin density 
is given by $m=n_{\uparrow}-n_{\downarrow}$.

The critical exponents which determine the response-function
properties and the associate instabilities of the many-electron 
problem $(1)$ are fully controlled by the pseudoparticle
interactions. Let us then provide some basic information
on the low-energy pseudoparticle operator basis which 
diagonalizes the quantum problem $(1)$. This diagonalization uses 
the BA \cite{Yang,Lieb}. We consider all finite values of $U$, 
electron densities $0<n<1$, and spin densities $0<m<n$. (In
Sec. V we also consider the $n=1$ half-filling and $m=n$
fully-polarized ferromagnetic phases.)
For this parameter space the low-energy physics is dominated 
by the lowest-weight states (LWS's) of the spin and $\eta$-spin algebras 
which in Refs. \cite{Carmelo94,Carmelo94b,Carmelo96} were called
of type I -- in this paper we call them for LWS's I. These are 
described by real BA rapidities, whereas all
or some of the BA rapidities which describe the LWS's II 
are complex and non-real \cite{Carmelo96}. Both the LWS's II and 
the non-LWS's out of the BA solution have energy gaps relative 
to each canonical ensemble ground state 
\cite{Carmelo94,Carmelo94b,Carmelo96} and do not contribute 
to the quantum problem instabilities studied in this paper.

In the Hilbert subspace spanned by the LWS's I the BA solution 
was shown to refer to an operator algebra which involves two types 
of {\it pseudoparticle} creation (annihilation) operators 
$b^{\dag }_{q,\alpha }$ ($b_{q,\alpha }$). These obey the 
usual anti-commuting algebra \cite{Carmelo94,Carmelo94b,Carmelo96} 

\begin{equation}
\{b^{\dag }_{q,\alpha},b_{q',\alpha'}\} 
=\delta_{q,q'}\delta_{\alpha ,\alpha'}, \hspace{0.5cm} 
\{b^{\dag }_{q,\alpha},b^{\dag }_{q',\alpha'}\}=0, \hspace{0.5cm} 
\{b_{q,\alpha},b_{q',\alpha'}\}=0 \, .
\end{equation}       
Here $\alpha$ refers to the two pseudoparticle colors $c$ and $s$ 
\cite{Carmelo94,Carmelo94b,Carmelo96}. The discrete pseudomomentum 
takes on values $q_j = {2\pi\over {N_a}}I_j^{\alpha }$, where in 
contrast to the usual momentum, $I_j^{\alpha }$ are  
consecutive integers or half-odd 
integers. It can be shown that in spite of the above
anti-commuting algebra the $\alpha $ pseudoparticles
have neither a fermionic nor a bosonic statistics \cite{Carmelo96c}. 
However, provided we introduce the topological-momentum shift 
operators studied in Refs. \cite{Carmelo96b,Carmelo96,Carmelo96c}
the second-quantization pseudoparticle representation
can be used. There are $d_{F\alpha }$ values of $I_j^{\alpha }$
($d_{F\alpha }$ is the $\alpha $-band Fock-space dimension), 
{\it i.e.} $j=1,...,d_{F\alpha }$. (The $\alpha $ 
topological-momentum shift operators produce shifts of $\pm {\pi\over N_a}$ 
in the $\alpha $-band pseudomomentum numbers $I_j^{\alpha }$ 
\cite{Carmelo96b,Carmelo96,Carmelo96c}.) A LWS I is specified by 
the distribution of $N_{\alpha }$ occupied values, which we call 
$\alpha $ pseudoparticles, over the $d_{F\alpha }$ available values. 
There are $d_{F\alpha }-N_{\alpha }$ corresponding empty values, 
which we call $\alpha $ pseudoholes \cite{Carmelo96}. These are 
good quantum numbers such that $d_{Fc}=N_a$, $N_c=N_{\uparrow}+ 
N_{\downarrow}$, $d_{Fs}=N_{\uparrow}$ and $N_s=N_{\downarrow}$.

The numbers $I_j^c$ are integers (or half-odd integers) for 
$N_s$ even (or odd), and $I_j^s$ are integers (or 
half-odd integers) for $d_{Fs}$ odd (or even) \cite{Lieb}. 
Therefore, all excitations which change $N_s=N_{\downarrow}$
and $d_{Fs}=N_{\uparrow}$ by an odd integer always involve 
topological-momentum shifts \cite{Carmelo96b,Carmelo96,Carmelo96c}.
All the LWS's I can be generated by acting onto
the vacuum $|V\rangle $ (zero-electron density) suitable
combinations of pseudoparticle operators 
\cite{Carmelo94,Carmelo94b} and are Slatter determinants of 
pseudoparticle levels of the simple form 

\begin{equation}
|\phi ; N_c, N_s\ \rangle = \prod_{\alpha=c,s}
[\prod_{q} b^{\dag }_{q,\alpha }] |V\rangle \, .
\end{equation}
We emphasize that all LWS's I are characterized by fixed values 
of $N_{s}$ and $d_{Fs}$ and therefore the pseudomomentum
takes on numbers $I_j^{\alpha }$ of well defined integer or
half-odd integer character. The ground state (GS) of a 
canonical ensemble with $N_c$ and $N_s$ pseudoparticle
numbers has the particular form

\begin{equation}
|GS; N_c, N_s\rangle = \prod_{\alpha=c,s}
[\prod_{q=q_{F\alpha }^{(-)}}^{q_{F\alpha }^{(+)}} 
b^{\dag }_{q,\alpha }] 
|V\rangle \, .
\end{equation}
When $N_{\alpha }$ is odd (even) and the numbers 
$I_j^{\alpha }$ are integers (half-odd integers) the pseudo-Fermi 
points are symmetric and given by \cite{Carmelo94,Carmelo94b}
$q_{F\alpha }^{(+)}=-q_{F\alpha }^{(-)} =
q_{F\alpha }-{\pi\over {N_a}}$ where

\begin{equation}
q_{F\alpha } = {\pi N_{\alpha}\over {N_a}} \, .
\end{equation}
On the other hand, when $N_{\alpha }$ is odd (even) and
$I_j^{\alpha }$ are half-odd integers (integers)
we have that $q_{F\alpha }^{(+)} = q_{F\alpha }$
and $-q_{F\alpha }^{(-)} =q_{F\alpha }-{2\pi\over {N_a}}$
or  $q_{F\alpha }^{(+)} = q_{F\alpha }-{2\pi\over {N_a}}$ 
and $-q_{F\alpha }^{(-)} = q_{F\alpha }$. 
Similar expressions are obtained for the pseudo-Brioullin 
zones limits $q_{c }^{(\pm)}$ if we replace in the above
expressions $N_{c }$ by $N_a$. On the other hand, the pseudo-Brioullin 
zones limits $q_{s }^{(\pm)}$ are always symmetric and
given by $q_{s }^{(+)}=-q_{s }^{(-)} =
{\pi\over {N_a}}[d_{Fs}-1]$.

In the pseudoparticle basis spanned by the LWS's I and 
in normal order relatively to the GS $(4)$ the 
Hamiltonian $(1)$ has the following form \cite{Carmelo94,Carmelo94b}

\begin{equation}
:\hat{H}: = \sum_{i=1}^{\infty}\hat{H}^{(i)} \, ,
\end{equation} 
where, to second pseudoparticle scattering order

\begin{eqnarray}
\hat{H}^{(1)} & = & \sum_{q,\alpha} 
\epsilon_{\alpha}(q):\hat{N}_{\alpha}(q): \, ;\nonumber\\
\hat{H}^{(2)} & = & {1\over {N_a}}\sum_{q,\alpha} \sum_{q',\alpha'} 
{1\over 2}f_{\alpha\alpha'}(q,q') 
:\hat{N}_{\alpha}(q)::\hat{N}_{\alpha'}(q'): \, .
\end{eqnarray} 
Here $(7)$ are the Hamiltonian terms which are {\it 
relevant} at low energy \cite{Carmelo94b}. 
Furthermore, at low energy and small momentum the only relevant 
term is the non-interacting term $\hat{H}^{(1)}$. Therefore,
the $c$ and $s$ pseudoparticles are non-interacting at the 
small-momentum and low-energy fixed point and the spectrum 
is described in terms of the bands $\epsilon_{\alpha}(q)$ 
studied in Ref. \cite{Carmelo91}.

At higher energies and (or ) large momenta the pseudoparticles 
start to interact via zero-momentum transfer forward-scattering 
processes of the Hamiltonian $(6)-(7)$. As in a Fermi liquid, 
these are associated with $f$ functions 
\cite{Carmelo92,Carmelo94} which read

\begin{eqnarray}
f_{\alpha\alpha'}(q,q') & = & 2\pi v_{\alpha}(q) 
\Phi_{\alpha\alpha'}(q,q')  
+ 2\pi v_{\alpha'}(q') \Phi_{\alpha'\alpha}(q',q) \nonumber \\
& + & \sum_{j=\pm 1} \sum_{\alpha'' =c,s}
2\pi v_{\alpha''} \Phi_{\alpha''\alpha}(jq_{F\alpha''},q)
\Phi_{\alpha''\alpha'}(jq_{F\alpha''},q') \, ,
\end{eqnarray}
where the pseudoparticle group velocities are given by 
$v_{\alpha}(q) = {d\epsilon_{\alpha}(q) \over {dq}}$
and $v_{\alpha }\equiv v_{\alpha }(q_{F\alpha})$
are the pseudo-Fermi group velocities. In expression
$(8)$ $\Phi_{\alpha\alpha '}(q,q')$ mesures the phase shift
of the $\alpha '$ pseudoparticle of pseudomomentum $q'$
due to the zero-momentum forward-scattering collision with the
$\alpha $ pseudoparticle of pseudomomentum $q$.
These phase shifts describe the pseudoparticle
interactions and are defined in Ref. \cite{Carmelo92}.
They control the low-energy physics. For instance, the 
related parameters

\begin{equation}
\xi_{\alpha\alpha '}^j = \delta_{\alpha\alpha '}+ 
\Phi_{\alpha\alpha '}(q_{F\alpha}^{(+)},q_{F\alpha '}^{(+)})+
(-1)^j\Phi_{\alpha\alpha '}(q_{F\alpha}^{(+)},q_{F\alpha '}^{(-)})
\, , \hspace{2cm} j=0,1 \, ,
\end{equation}
play a determining role in the expressions we evaluate
in this paper. There is a simple and direct relation
between the pseudoparticle-interaction parameters and
the quantities of CFT. For instance, the anti-symmetric combinations
of two-pseudoparticle phase shifts $(9)$, $\xi_{\alpha\alpha 
'}^1$, are nothing but the entries of the transpose of the 
dressed-charge matrix \cite{Frahm}. Both the symmetric
[$j=0$ in Eq. $(9)$] and anti-symmetric [$j=1$ in Eq. $(9)$]
combinations of two-pseudoparticle phase shifts $(9)$ fully 
control the instabilities of the quantum problem, as we find 
in the next sections.

\section{THE INSTABILITIES OF THE QUANTUM PROBLEM}

In order to study the divergences of the low-frequency 
response functions $\hbox{Re}\chi^{\vartheta } (k,\omega)$ 
we start by characterizing the relevant GS transitions 
which determine these divergences.

Following Ref. \cite{Carmelo94b}, at low energy there are
independent right $N_{\alpha}^+$ and left $N_{\alpha}^-$ 
$\alpha $ pseudoparticle conservation laws in the Hubbard chain.
Here $N_{\alpha}^+$ (or $N_{\alpha}^-$) are the number
of $\alpha $ pseudoparticles, with $q>0$ (or $q<0$).
While fixed values of $N_c$ and $N_s$ (and then of
$N_{\uparrow}$ and $N_{\downarrow}$ electron numbers)
define a canonical ensemble, we introduce the concept
of {\it sub canonical ensemble} which refers to fixed
values of $N_c^+$, $N_c^-$, $N_s^+$, and $N_s^-$.
Obviously, the same canonical ensemble is realized
by several sub-canonical ensembles. 

Since all LWS's I of form $(3)$ are characterized by
fixed values of the pseudoparticle numbers $N_c^+$, $N_c^-$, 
$N_s^+$, and $N_s^-$, we can associate with each
sub-canonical ensemble a low-energy Hilbert subspace 
spanned by these LWS's I which have the same pseudoparticle
numbers. We call pseudo-ground state (PGS) the
LWS I (or LWS's I) of minimal energy in each of
these Hilbert sub spaces. All PGS's are of the
form

\begin{equation}
|PGS; D_c, D_s\rangle = \prod_{\alpha=c,s}
[\prod_{q=q_{F\alpha }^{(-)}+D_{\alpha}{2\pi\over 
{N_a}}}^{q_{F\alpha }^{(+)}+D_{\alpha}{2\pi\over {N_a}}} 
b^{\dag }_{q,\alpha }] 
|V\rangle \, ,
\end{equation}
where

\begin{equation}
D_{\alpha} = \Delta N_{\alpha}^+ = -\Delta N_{\alpha}^-
\end{equation}
refers to the $\Delta N_{\alpha}^{\pm}$ changes relative
to the GS $(4)$ with the same $N_c$ and $N_s$ numbers.

An important point is that the transitions between a GS
and any LWS I can be separated into two types of 
excitations: (a) a GS - GS or GS - PGS transition
which involves changes in the pseudoparticle numbers and (b) a
Landau-liquid excitation associated with pseudoparticle-pseudohole
processes relative to the final GS or PGS which do
not change the right and left pseudoparticle numbers.

A remarkable property is that the low-frequency 
correlation-function divergences are fully determined
by the above transitions (a), followed by transitions (b). 
This is because these divergences are determined by the
quantum overlap between GS's and PGS's and the
correlation-function excitations. The momentum of
these transitions (a) is given by

\begin{equation}
k_0 = \sum_{\alpha} D_{\alpha} 2q_{F\alpha} \, ,
\end{equation}
where for the pure GS - PGS transitions (by pure GS - PGS
transitions we mean here those which do not change the
$N_c$ and $N_s$ pseudoparticle numbers) $D_{\alpha}=0, \pm 1, 
\pm 2, ...$ are always integers, whereas for the GS - GS
transitions $D_{\alpha}$ can either be integers or
half-odd integers.

The pseudoparticle-pseudohole processes (b) increase
the value of the critical exponents which control
the response-function divergences by positive integers.
This leads to positive exponents associated
with irrelevant correlation-function terms.
Therefore, the divergences occur always precisely at 
momentum values $k=k_0$ (see Eq. $(12)$) which 
correspond to GS - GS or GS - PGS transitions (a).

In order to find the divergences of {\it all} low-frequency 
response functions $\hbox{Re}\chi^{\vartheta } (k,\omega)$ 
we have followed the steps of Refs. \cite{Carmelo93,Carmelo96b} 
and combined our suitable generator pseudoparticle analysis 
\cite{Carmelo94,Carmelo94b} with the results obtained from
CFT \cite{Frahm,Frahm91}.
The asymptotic expression of the correlation function 
$\chi^{\vartheta } (x,t)$ in $x$ and $t$ 
space is given by the summation of many terms of form $(3.13)$ 
of Ref. \cite{Frahm} with dimensions of the fields suitable 
to that function. For small energy the corresponding correlation 
functions in $k$ and $\omega $ space are obtained by the summation 
of the Fourier transforms of these terms, which are of the form 
given by Eq. $(5.2)$ of Ref. \cite{Frahm91}. However, the results of 
Refs. \cite{Frahm,Frahm91} do not provide the specific expression 
at $k=k_0$ and small positive $\omega $. In this case the above 
summation is equivalent to a summation in the suitable final 
GS's or PGS's which correspond to different values for the 
dimensions of the fields. 

As was already referred in Sec. I, we emphasize that expression 
$(5.7)$ of Ref. \cite{Frahm91} {\it is not} valid in our case. 
While we need the general response-function $\omega $ dependence
at $k=k_0$, expression $(5.7)$ of Ref. \cite{Frahm91} is only 
valid for a very particular relation between $(k-k_0)$ and
$\omega$. Moreover, the limit $(k-k_0)\rightarrow 0$ of
that expression {\it does not} provide the general low-$\omega $
behavior at $k=k_0$.

Following Ref. \cite{Carmelo96b}, we have solved the following 
general integral 

\begin{equation}
\tilde{g}(k_0,\omega ) = \int_{0}^{\infty}dx\int_{-\infty}^{\infty}
dt e^{i\omega t} \prod_{\alpha ,\iota}
{1\over {(x+\iota v_{\alpha }t)^{2\Delta_{\alpha }^{\iota}}}} 
\, ,  
\end{equation}
which is the integral of expression $(5.2)$ of Ref. \cite{Frahm91}
at $k=k_0$. In Eq. $(13)$ $\Delta_{\alpha }^{\iota}$ with $\iota =+1$ 
and $\iota =-1$ for right and left $\alpha $ pseudoparticles, 
respectively, are the GS - GS or GS - PGS suitable dimensions of 
the fields \cite{Frahm,Carmelo94} of the general form

\begin{equation}
2\Delta_{\alpha }^{\iota}\equiv 
\left[\sum_{\alpha '}\xi_{\alpha\alpha '}^1 D_{\alpha '}
+ \iota\sum_{\alpha '}\xi_{\alpha\alpha '}^0 {\Delta N_{\alpha 
'}\over 2}\right]^2 \, , 
\end{equation}
where $\xi_{\alpha\alpha '}^j$ (with $j=0,1$) are the
combinations of two-pseudoparticle phase shifts $(9)$
and $\Delta N_{\alpha}$ refers to the changes in the
numbers $N_{\alpha}$ associated with the transition.
We find for the integral $(13)$, $\tilde{g}(k_0,\omega )\propto
\omega^{\varsigma_{\vartheta}}$. Here  

\begin{equation}
\varsigma_{\vartheta} =
2\sum_{\alpha ,\iota}\Delta_{\alpha }^{\iota} - 2 \, .
\end{equation}
Comparing our expression with expression $(5.7)$ of Ref.  \cite{Frahm91} we 
confirm these expressions are different. This is because these
expressions correspond to different limiting cases which
do not commute. We find that at the momenta $k=k_0$ 
[given by Eq. $(12)$] and low frequency $\omega $ all correlation functions 
behave as $\hbox{Im}\chi^{\vartheta}(k_0,\omega)\propto 
\omega^{\varsigma_{\vartheta }}$ with the corresponding
real part given by

\begin{equation}
\hbox{Re}\chi^{\vartheta}(k_0,\omega)\propto 
\omega^{\varsigma_{\vartheta }} \, ,
\end{equation}
for $\varsigma_{\vartheta }\neq 0$ and
by $\hbox{Re}\chi^{\vartheta}(k_0,\omega)\propto
-\ln (\omega )$ for $\varsigma_{\vartheta }=0$,
respectively. Given this related expressions for the imaginary and
real parts, below we consider only the latter part.
Our task is detecting all response functions and
momenta values $k=k_0$ for which the corresponding
exponent $\varsigma_{\vartheta }$, Eq. $(15)$, is zero or
negative. We have evaluated it for all correlation
functions and possible values of $k_0$. All divergences
occur only for some of the correlation functions which show
logarithmic divergences at $U=0$, ie all divergences occurring
at finite values of $U$ correspond to logarithmic divergences
at $U=0$. However, we find that the onsite electron-electron 
interactions supress such divergences in the case of the the 
superconductivity functions and for
some values of $U$ in the case of the charge and spin
functions $\hbox{Re}\chi^{\rho} (\pm 2k_{F\uparrow},\omega)$
and $\hbox{Re}\chi^{\sigma_z} (\pm 2k_{F\uparrow},\omega)$,
respectively. In general the effects of electron correlations replace
the logarithmic divergence by a Luttinger-liquid \cite{Haldane}
power-law divergence.

The use in Eqs. $(15)$ and $(16)$ of suitable dimensions of the
fields $(14)$ for the different correlation functions leads
to the following results. For the transverse-spin function 
we find

\begin{equation}
\hbox{Re}\chi^{\sigma_{\perp}}(\pm 2k_F,\omega)\propto 
\omega^{\varsigma_{s\perp }} \, , 
\end{equation}
where the exponent is such that $-1<\varsigma_{s\perp}<0$ and 
given by

\begin{equation}
\varsigma_{s\perp} = -2 + 2\sum_{\alpha}
[({\xi_{\alpha c}^1\over 2})^2+({\xi_{\alpha s}^0\over 2})^2] \, .
\end{equation}
The divergences of the charge and spin functions were 
already studied in Ref. \cite{Carmelo93} and are
controlled by the same exponents. The expressions read

\begin{equation}
\hbox{Re}\chi^{\rho} (\pm 2k_{F\sigma},\omega) 
\propto \omega^{\varsigma_{cs\sigma}} \, ,  \hspace{1cm}
\hbox{Re}\chi^{\sigma_z } (\pm 2k_{F\sigma},\omega) 
\propto \omega^{\varsigma_{cs\sigma}} \, , 
\end{equation}
where the exponents are such that
$-{1\over 2}<\varsigma_{cs\downarrow}<0$ 
and $-{1\over 2}<\varsigma_{cs\uparrow}<2$ 
and are given by

\begin{equation}
\varsigma_{cs\downarrow} = -2 + 2\sum_{\alpha}
(\xi_{\alpha s}^1)^2 \, , \hspace{1cm}
\varsigma_{cs\uparrow} = -2 + 2\sum_{\alpha}
(\xi_{\alpha c}^1 - \xi_{\alpha s}^1)^2 \, .
\end{equation}
(We emphasize that at the parameter-space points which
correspond to $\varsigma_{cs\uparrow}=0$ the expressions
of Eq. $(19)$ are not valid and should be replaced by
logarithmic $\omega $ dependences.)

In the case of the singlet-superconductivity function we 
find

\begin{equation}
\hbox{Re}\chi^{ss} (\pm 2k_F,\omega ) 
\propto \omega^{\varsigma_{ss+}} \, , \hspace{1cm}
\hbox{Re}\chi^{ss} (\pm [k_{F\uparrow} - k_{F\downarrow}],\omega)
\propto \omega^{\varsigma_{ss-}} \, , 
\end{equation}
where the exponents are such that $0<\varsigma_{ss+}<1$ 
and $0<\varsigma_{ss-}<3$ and are given by

\begin{eqnarray}
\varsigma_{ss+} & = & -2 + 2\sum_{\alpha}
[({\xi_{\alpha c}^1\over 2})^2+(\xi_{\alpha c}^0+
{\xi_{\alpha s}^0\over 2})^2] \nonumber \\
\varsigma_{ss-} & = & -2 + 2\sum_{\alpha}
[({\xi_{\alpha c}^1\over 2} - \xi_{\alpha s}^1)^2+(\xi_{\alpha c}^0 
+ {\xi_{\alpha s}^0\over 2})^2] \, .
\end{eqnarray}
Since for $U>0$ these exponents are positive (as the figures
of next section confirm) it 
follows that the electron-electron interactions remove the
$U=0$ singlet-superconductivity instability. The same occurs
with the triplet $\sigma $ superconductivity function which 
reads

\begin{equation}
\hbox{Re}\chi^{ts\sigma} (0,\omega)
\propto \omega^{\varsigma_{t\sigma s}} \, , 
\end{equation}
where the exponent is such that $0<\varsigma_{ts\downarrow}<2$ 
and $0<\varsigma_{ts\uparrow}<1$ and is given by

\begin{equation}
\varsigma_{ts\downarrow} = -2 + 2\sum_{\alpha}
(\xi_{\alpha c}^0 + \xi_{\alpha s}^0)^2 \, , \hspace{1cm}
\varsigma_{ts\uparrow} = -2 + 2\sum_{\alpha}
(\xi_{\alpha c}^0)^2 \, .
\end{equation}

The $\sigma $ one-electron Green function was already 
studied in Ref. \cite{Carmelo96b} and for small $\omega$ and 
$U>0$ is such that

\begin{equation}
\hbox{Re}G_{\sigma}(\pm k_{F\sigma},\omega)\propto 
\omega^{\varsigma_{\sigma}} \, , 
\end{equation}
where the exponent $-1<\varsigma_{\sigma} <-1/2$
is given by

\begin{equation}
\varsigma_{\uparrow}=-2+\sum_{\alpha}
{1\over 2}[(\xi_{\alpha c}^1-\xi_{\alpha s}^1)^2
+(\xi_{\alpha c}^0)^2] \, ,  \hspace{1cm}
\varsigma_{\downarrow}=-2 
+\sum_{\alpha}{1\over 2}[(\xi_{\alpha s}^1)^2+(\xi_{\alpha 
c}^0+\xi_{\alpha s}^0)^2] \, .
\end{equation}
 
All exponents $(18)$, $(20)$, $(22)$, $(24)$, and $(26)$ have 
a Luttinger-liquid non-classical character being $U$, $n$, and $m$ 
dependent functions. They are fully controlled by the two-pseudoparticle 
forward-scattering collisions [see Eqs. $(9)$ and $(16)$].

\section{COMPARATIVE STUDY OF THE INSTABILITIES}

In this section we present and discuss the $U$, electronic 
density, and magnetic-field dependence of the exponents
$(18)$, $(20)$, $(22)$, $(24)$, and $(26)$ which control the 
quantum-problem instabilities. Our comparative study of these 
exponents allows finding the dominant instabilities in the 
different regions of parameter space.

In the Table we present various limiting values of the 
correlation-function exponents. There the $m\rightarrow n$ values 
are also valid when $H\rightarrow H_c$, $U\rightarrow U_c$, and 
$n\rightarrow n_c$, where $H_c$, $U_c$, and $n_c$ are the 
(interrelated) critical values for the onset of ``ferromagnetism'' 
({\it i.e.}, full spin polarization, $m\rightarrow n$)
and are defined by Eq. $(2)$ of Ref. \cite{Carmelo92c}. 
Below we often use the normalized magnetic field $h=H/H_c$.
In the limit $n\rightarrow 1$ (not presented in the 
Table) the exponent expressions also simplify 
because $\xi_{cc}^1=\xi_{cc}^0=1$ and 
$\xi_{sc}^1=\xi_{cs}^0=0$ and the remaining
parameters change from $\xi_{cs}^1=1/2$, $\xi_{ss}^1=1/\sqrt{2}$,
$\xi_{sc}^0=-1/\sqrt{2}$, and $\xi_{ss}^0=\sqrt{2}$ 
when $m\rightarrow 0$ to $\xi_{cs}^1=0$, $\xi_{ss}^1=1$,
$\xi_{sc}^0=0$, and $\xi_{ss}^0=1$ as $m\rightarrow n$.

At finite values of $U$ the exponents $\varsigma_{\uparrow}$, 
$\varsigma_{\downarrow}$, $\varsigma_{s\perp}$, and 
$\varsigma_{cs\downarrow}$ are always negative, whereas
$\varsigma_{cs\uparrow}$ is only negative for some
regions of parameter space. This exponent is negative
in the limit of $m\rightarrow 0$ for all values
of the density and $U$. For $0<m<n$ it
can have both negative and positive values \cite{Carmelo93}. 
It is an increasing
function of the magnetic field, becoming positive
for a value of that field which depends on $n$ and $U$.
In the limit $m\rightarrow n$ it becomes positive for all
finite values of $U$ and all electronic densities $n$.
For finite values of $U$ all the remaining exponents are 
positive. In the figures we use often the electronic
density $n=0.5$ because it is typical for many of the
quasi-one-dimensional Beckgaard salts \cite{Ishiguro,Gruner}.

In Fig. 1 the exponents
(a) $\varsigma_{s\perp}$, (b) $\varsigma_{cs\downarrow}$ and
$\varsigma_{cs\uparrow}$, (c) $\varsigma_{\downarrow}$,
and (d) $\varsigma_{\uparrow}$ and in Fig. 2 the superconductivity
exponents (a) $\varsigma_{ss+}$, (b) $\varsigma_{ss-}$,
(c) $\varsigma_{ts\downarrow}$, and (d) $\varsigma_{ts\uparrow}$
are plotted vs $U$ for $n=0.5$ and various values
the of the magnetic field.

While the exponents $\varsigma_{s\perp}$ and $\varsigma_{cs\downarrow}$ 
are always a decreasing function of $U$ and the exponent 
$\varsigma_{cs\uparrow}$ a deacreasing and increasing function
of $U$ for smaller and large values of $h$, respectively,
all remaining above exponents are always an increasing function
of $U$.

In Fig. 3 the exponents (a) $\varsigma_{s\perp}$, (b) 
$\varsigma_{cs\downarrow}$ and $\varsigma_{cs\uparrow}$, 
(c) $\varsigma_{\downarrow}$, and (d) $\varsigma_{\uparrow}$ 
and in Fig. 4 the superconductivity
exponents (a) $\varsigma_{ss+}$, (b) $\varsigma_{ss-}$,
(c) $\varsigma_{ts\downarrow}$, and (d) $\varsigma_{ts\uparrow}$
are plotted vs the electronic density for $U=10$
and three values of the magnetic field.

All the exponents are smooth functions of the density showing
either a maximum or a minimum for an intermediate value
of the density. While the exponents $\varsigma_{s\perp}$
and $\varsigma_{cs\sigma}$ show a maximum value, the
Green-function and superconductivity exponents show
a minimum value.

The magnetic-field dependences of the correlation-function 
exponents are illustrated in Fig. 5 and 6
where these parameters are plotted as functions of 
$h$ for density $n=0.5$ and various values of $U$. 
While the exponents $\varsigma_{s\perp}$ and
$\varsigma_{ts\uparrow}$ are always a decreasing function
of the magnetic field, all the remaining exponents
plotted in the figures are in general an increasing
function of $h$.

We emphasize that the exponent $\varsigma_{\uparrow}$
is always smaller than or equal to the exponent
$\varsigma_{\downarrow}$. 
In order to understand the parameter-space critical line
defined by the equation $\varsigma_{s\perp}=\varsigma_{\uparrow}$,
the exponents $\varsigma_{s\perp}$ and $\varsigma_{\uparrow}$
associated with the two dominant instabilities are plotted
in Figs. 7 (a)-(d) as functions of the magnetic field $h$ and 
for several values of the electronic density $n$ and onsite
interaction (a) $U=1$, (b) $U=3$, (c) $U=5$, and (d)
$U=20$. At zero magnetic field $\varsigma_{\downarrow}
=\varsigma_{\uparrow}$, $\varsigma_{\uparrow} 
<\varsigma_{s\perp}$, and the exponent $\varsigma_{s\perp}$
equals the exponents $\varsigma_{cs\downarrow}$ and
$\varsigma_{cs\uparrow}$ which are equal in that limit. 
(The latter were already studied in Ref. \cite{Carmelo93}.) 
Therefore, adding or removing of electrons is in this case 
the dominant instability. (In the case of the weakly
coupled-chain Hubbard system this leads to a dominance
for interchain electron hopping.) Also at small values
of $U$ that instability remains dominant for all values
of the electronic density and magnetic field.

However, the magnetic field removes the $SU(2)$ spin
symmetry and for intermediate and large $U$ values induces a 
dominant character for the spin-flip excitation over adding or 
removing of single electrons. This occurs for $U>U_c^*$,
where $U_c^*=U_c^*(n,h)$ is a critical value. Depending on the 
parameters that we keep constant, we can also define critical 
values $n_c^*(U,h)$ and $h_c^*(n,U)$. The critical values
are obtained by solution of the above equation,
$\varsigma_{\uparrow}=\varsigma_{s\perp}$,
upon suitable boundary conditions. Therefore, they
correspond to the points of Figs. 7 (a)-(d) where
the two exponent lines cross. As these figures confirm, 
$U_c^*$ changes from 

\begin{equation}
U_c^*= {4t\sin (\pi n)\over {\tan ({\pi\over 2}
[\sqrt{2}-1])}} \, ,
\end{equation}
when $h\rightarrow 1$ to $U_c^*=\infty$ when $h\rightarrow 0$. 

In Figs. 8 and 9 the functions $U_c^*(n)$ and 
$h_c^*(n)$ are presented for constant values of $h$ and
$U$, respectively. These define the critical lines which
divide the regions of parameter space where $\uparrow $-electron
adding and removing on the one hand and spin-transverse
SDW on the other hand are the dominant instability.

At finite values of the magnetic field $h<1$, $\uparrow $-electron 
adding or removing remains dominant for $U<U_c^*$ (see Fig. 8). 
This region is larger for intermediate densities 
and for small values of the magnetic field and smaller for small 
densities and densities close to half filling and for large 
values of $h$. Therefore, the function $U_c^*(n)$ plotted in Fig. 8
shows a maximum at intermediate electronic densities and
is larger at smaller values of the magnetic field $h$.

On the other hand, at intermediate and large values of
$U$, $\uparrow $-electron adding or removing remains
dominant for $h<h_c^*$ (see Fig. 9). This region is
larger for intermediate densities and for small values
of $U$ and smaller for small densities and densities close
to $n=1$ and for large values of $U$. It follows that
the function $h_c^*(n)$ plotted in Fig. 9
shows a maximum for intermediate values of the electronic
density and is larger for smaller values of $U$.

\section{THE METAL - INSULATOR HALF-FILLING AND FULLY-POLARIZED 
	 FERROMAGNETIC TRANSITIONS}

In this section we consider only these correlation functions
whose exponents are negative for finite $U$ and omit the
study of the superconductivity correlation functions.

When we change the chemical potential $\mu$ from 
$\mu>{\Delta_{MH}\over 2}$ to $\mu<{\Delta_{MH}\over 2}$, where 
$\Delta_{MH}$ is the half-filling 
Mott-Hubbard gap \cite{Lieb} which at finite magnetic field was 
studied in Ref. \cite{Carmelo91} (and thus the electronic density 
changes from $n<1$ to $n=1$), there occurs in the many-electron
system a metal - insulator transition \cite{Carmelo92c}.
Also when we change the magnetic field from $H<H_c$ to $H>H_c$ 
(and thus the spin density $m$ changes from $m<n$ to $m=n$), there 
occurs a transition to fully-polarized
ferromagnetism. In this section we find that i)- some
exponents show a discontinuity because they have different
values in both sides of the above
transitions and ii)- these transitions remove some
of the low-energy power-law divergences in the correlation 
functions but in some cases these divergences are shifted to 
finite energies.

At half filling both the charge and the one-particle 
excitations associated with creation of one electron cost a
minimal energy $\Delta_{MH}$. Therefore, both the
charge response function and the one-electron
Green function vanish for positive energy, $\omega <\Delta_{MH}$.

On the other hand, there are gapless spin excitations which lead to
low-energy power-law behavior for the spin response
functions. However, the half-filling metal - insulator transition
leads to discontinuities in the corresponding exponents.
We find

\begin{equation}
\hbox{Re}\chi^{\sigma_{\perp}}(\pm\pi,\omega)\propto 
\omega^{\varsigma_{s\perp }} \, , \hspace{1cm}
\hbox{Re}\chi^{\sigma_z } (\pm 2k_{F\sigma},\omega) 
\propto \omega^{\varsigma_{cs\sigma}} \, , 
\end{equation}
where the exponents $\varsigma_{s\perp}$, 
$\varsigma_{cs\downarrow}$, and $\varsigma_{cs\uparrow}$ 
are for $n\rightarrow 1$ and {\it at} $n=1$ 
(and for $\mu\rightarrow {\Delta_{MH}\over 2}$ from above and
for $\mu<{\Delta_{MH}\over 2}$) given by

\begin{equation}
\varsigma_{s\perp} = -2[{3\over 4}  
- ({1\over 2\xi_{ss}^1})^2] \, , \hspace{0.5cm}
\varsigma_{cs\downarrow} = -2[1-(\xi_{cs}^1)^2 
-(\xi_{ss}^1)^2] 
\, , \hspace{0.5cm}
\varsigma_{cs\uparrow} = -2[1 - (1 - \xi_{cs}^1)^2 
- (\xi_{ss}^1)^2] 
\, ,
\end{equation}
and

\begin{equation}
\varsigma_{s\perp} = -2[1 - ({1\over 2\xi_{ss}^1})^2] \, , \hspace{1cm}
\varsigma_{cs\downarrow} = -2[1 - (\xi_{ss}^1)^2] 
\, , \hspace{1cm}
\varsigma_{cs\uparrow} = -2[1 -
(\xi_{ss}^1)^2] 
\, ,
\end{equation}
respectively. In the limit $m\rightarrow 0$ these exponents 
read

\begin{equation}
\varsigma_{s\perp} = \varsigma_{cs\downarrow} = 
\varsigma_{cs\uparrow} = -1/2 \, ,
\end{equation}
for $n\rightarrow 1$ (and for $\mu\rightarrow {\Delta_{MH}\over 2}$
from above) and

\begin{equation}
\varsigma_{s\perp} = \varsigma_{cs\downarrow} = 
\varsigma_{cs\uparrow} = -1 \, ,
\end{equation}
for $n=1$ (and for $\mu<{\Delta_{MH}\over 2}$). 

On the other hand, in the limit $m\rightarrow n$ these exponents 
are given by

\begin{equation}
\varsigma_{s\perp} = -1 \, , \hspace{1cm}
\varsigma_{cs\downarrow} = 0
\, , \hspace{1cm}
\varsigma_{cs\uparrow} = 2 \, ,
\end{equation}
for $n\rightarrow 1$ (and for $\mu\rightarrow {\Delta_{MH}\over 2}$
from above) and

\begin{equation}
\varsigma_{s\perp} = -3/2 \, , \hspace{1cm}
\varsigma_{cs\downarrow} = 0 
\, , \hspace{1cm}
\varsigma_{cs\uparrow} = 0
\, ,
\end{equation}
for $n=1$ (and for $\mu<{\Delta_{MH}\over 2}$). 
Obviously, in the cases where the exponents vanish
expressions $(28)$ should be replaced by a logarithmic function.
Note that the logarithmic behavior is only reached in
the limit $m\rightarrow n$ since otherwise the correspondent
exponents are different from zero.

In the case of fully-polarized ferromagnetism there are no
spin-down electrons in the system. In this case a spin flip costs
a minimal energy given by

\begin{equation}
\Delta_F =2\mu_0[H-H_c] \, ,
\end{equation} 
where $H>H_c$ and $\Delta_F$ is the fully-polarized ferromagnetism 
gap. Therefore, the transverse-spin response
function and the down-spin one-electron Green function vanish for 
positive energy, $\omega <\Delta_F$. Also, there is no 
$k=k_{F\downarrow}=0$ divergence in
$\hbox{Re}\chi^{\rho} (k,\omega)$ and 
$\hbox{Re}\chi^{\sigma_z } (k,\omega)$. 

The charge response function, 
spin response function $\hbox{Re}\chi^{\sigma_z }$, and
up-spin one-electron Green function all have a
low-energy power-law structure in both sides of the
ferromagnetic transition. That transition leads to discontinuities 
in the exponents of these correlation functions, which
read

\begin{equation}
\hbox{Re}\chi^{\rho} (\pm 2k_F,\omega) 
\propto \omega^{\varsigma_{cs\uparrow}} 
\, , \hspace{1cm}
\hbox{Re}\chi^{\sigma_z } (\pm 2k_F,\omega) 
\propto \omega^{\varsigma_{cs\uparrow}} \, , \hspace{1cm}
\hbox{Re}G_{\uparrow}(\pm 2k_F,\omega)\propto 
\omega^{\varsigma_{\uparrow}} \, , 
\end{equation}
where for $m\rightarrow n$ (and $H\rightarrow H_c$ from
below) the exponents are given by

\begin{equation}
\varsigma_{cs\uparrow} = 2[1-\eta_0]^2 \, , \hspace{1cm}
\varsigma_{\uparrow} = -1+{1\over 2}[1-\eta_0]^2 \, ,  
\end{equation}
and

\begin{equation}
\eta_0 =({2\over {\pi}})
\tan^{-1}\left({4t\sin (\pi n)\over U}\right) \, .
\end{equation}

On the other hand, at $m=n$ (and $H>H_c$) the above charge 
and spin response functions have a non-interacting 
logarithmic behavior 

\begin{equation}
\hbox{Re}\chi^{\rho} (\pm 2k_F,\omega) 
\propto -\ln (\omega )
\, , \hspace{1cm}
\hbox{Re}\chi^{\sigma_z } (\pm 2k_F,\omega) 
\propto -\ln (\omega ) \, , 
\end{equation}
because

\begin{equation}
\varsigma_{cs\uparrow} = 0 \, ,
\end{equation}
and the up-spin Green function exponent reads

\begin{equation}
\varsigma_{\uparrow} = -1 \, .  
\end{equation}

Based on the occurrence of only zero-momentum pseudoparticle forward
scattering and on the pseudoparticle-number conservation
laws \cite{Carmelo96c} we find that in the case of half filling 
there are power-law divergences at finite energies, for instance, in 
the one-electron Green function. Finite-energy power-law 
divergences also occur, for example, in the transverse-spin response
function in the fully-polarized ferromagnetic phase. (The above property of the 
pseudoparticle interactions is valid at all energy scales 
\cite{Carmelo96c}. A complete study of the quantum problem for 
all energies involves an infinite number of pseudoparticle 
branches -- fortunately, the finite-energy divergences we study 
in this paper only involve the $c$ and $s$ branches.)

By generalizing the pseudoparticle operator basis to half filling
and the fully-polarized phase, we find an equivalent
low-energy pseudoparticle theory where some of the low-energy $\omega $
divergences of Sec. III are replaced by low-energy $(\omega 
-\Delta_{MH})$ and $(\omega -\Delta_{F})$, respectively, 
divergences. At these energies the half-filling
and fully-polarized ferromagnetic pseudoparticle theories involve
the same parameters as the general $0<n<1$ problem studied in
previous sections. For instance, for $(\omega -\Delta_{MH})>0$
and $(\omega -\Delta_F)>0$ all pseudoparticle-Hamiltonian
parameters are obtained by considering the limits $n\rightarrow 1$ 
and $m\rightarrow n$, respectively, of the corresponding $0<n<1$ 
and $0<m<n$ expressions (and not the corresponding $n=1$ and $m=n$ 
expressions).

In the case of half filling the lack of charge gapless excitations 
and the associate opening of the Mott-Hubbard gap leads to the following 
expression for the one-electron Green function

\begin{equation}
\hbox{Re}G_{\sigma}(\pm k_{F\sigma},\omega)\propto 
[\omega - \Delta_{MH}]^{\varsigma_{\sigma}} \, , 
\end{equation}
where the exponent $\varsigma_{\sigma}$ is given by

\begin{equation}
\varsigma_{\uparrow}=-2+{1\over 2}[(1-\xi_{cs}^1)^2 + 1 +
(\xi_{ss}^1)^2+(\xi_{sc}^0)^2] \, ,  
\end{equation}

\begin{equation}
\varsigma_{\downarrow}=-2+{1\over 2}[
(\xi_{cs}^1)^2+1+(\xi_{ss}^1)^2+
(\xi_{sc}^0+\xi_{ss}^0)^2] \, ,
\end{equation}
and $(\omega -\Delta_{MH})$ is small and positive. 

Both in the limits $m\rightarrow 0$ and $m\rightarrow n$ the 
exponents $(43)$ and $(44)$ for up-spin and down-spin
electrons, respectively, are equal and read     

\begin{equation}
\varsigma_{\uparrow} = \varsigma_{\downarrow} = -7/8 \, , 
\end{equation}
and

\begin{equation}
\varsigma_{\uparrow} = \varsigma_{\downarrow} = -1/2 \, ,
\end{equation}
respectively.

On the other hand, in the fully-polarized ferromagnetic phase
there are no gapless excitations in the transverse-spin 
channel. We find the following expression

\begin{equation}
\hbox{Re}\chi^{\sigma_{\perp}}(\pm 2k_F,\omega)\propto 
[\omega -\Delta_F ]^{\varsigma_{s\perp }} \, ,
\end{equation}
where the exponent is given by

\begin{equation}
\varsigma_{s\perp}=-1+(\eta_0)^2 \, .
\end{equation}
Obviously, this expression refers to
small positive values of the energy $(\omega -\Delta_F)$
and is not valid in the limit of $U\rightarrow 0$ where
$\varsigma_{s\perp}\rightarrow 0$.

Equations $(42)-(48)$ reveal that the opening of the
Mott-Hubbard and ferromagnetic gaps associated with
the electronic-correlation and magnetic effects, respectively,
moves some of the $\omega\rightarrow 0$
instabilities to $\omega\rightarrow\Delta_{MH}$ and
$\omega\rightarrow\Delta_F$, respectively. These
divergences occur at the same momenta values than
the corresponding low-frequency $0<n<1$ and
$0<m<n$ divergences.

The finite-frequency expressions $(42)-(48)$ cannot
be derived by CFT. Their validity follows from the purely
non dissipative character of the pseudoparticle interactions
which survives at all energy scales of the quantum problem
\cite{Carmelo96c}.

\section{COUPLED CHAINS}

The exponents calculated in the preceeding sections give the scaling
dimension of the corresponding operators, in the presence of the 
interaction. Once they are known, we can analyze the effects of 
perturbations added to (\ref{hamil}) which can be written as a sum
of terms containing the operators analyzed above. For simplicity, we
first examine the case in which these operators couple two different
chains. Let us write $\chi^{\vartheta} ( x , t ) = \langle 
{\cal O}^{\vartheta} ( x , t ) {\cal O}^{\vartheta} \rangle$.  
The Fourier thansform of $\chi_{\vartheta}$ goes at low frequencies as
$\omega^{\varsigma_{\vartheta}}$. An interchain  coupling term 
which involves the operator ${\cal O}^{\vartheta}$ can be written as

\begin{equation}
{\cal H}_{1 , 2} =  J_{\vartheta} \int dx {\cal O}_1^{\vartheta} ( x ) 
{\cal O}_2^{\vartheta} ( x ) \, ,
\label{coupling}
\end{equation} 
where the labels 1 and 2 refer to the two different chains.
We now add this term to (\ref{hamil}) and study the scaling of 
$J_{\vartheta}$ as the cutoff is reduced. We define the dimensionless
coupling ${\tilde{J}}_{\vartheta} = J_{\vartheta} / \omega_c$,
where $\omega_c$ is an upper cutoff below for which the approximation of
$\chi_{\vartheta}$ by a power law is approximately valid.
Then, neglecting the influence of the other interchain couplings 
we find\cite{Wen,Schulz1,Fabrizio}

\begin{equation}
\omega_c \frac{\partial {\tilde{J}}_{\vartheta}}{\partial \omega_c}
=  ( - 1 - 2 \varsigma_{\vartheta} )  {\tilde{J}}_{\vartheta} \, .
\label{scaling}
\end{equation}
Thus, the relevance of ${\cal O}_{\vartheta}$ depends on whether
$\varsigma_{\vartheta}$ is greater or less than $-1/2$.
When $\varsigma_{\vartheta} < - 1 / 2$, ${\cal O}_{\vartheta}$ is a
relevant operator which grows upon scaling. The Renormalization-Group
scheme used to derive (\ref{coupling}) breaks down when 
${\tilde{J}}_{\vartheta} \sim 1$.  This defines a crossover 
scale,

\begin{equation}
\epsilon_{\vartheta} = J_{\vartheta} \left( \frac{J_{\vartheta}}
{\omega_c} \right)^{- \frac{1}{1 + 2 \varsigma_{\vartheta}}} \, .
\label{crossover}
\end{equation}
Below this crossover the coupling $J_{\vartheta}$ dominates the physics
of the coupled-chain system and a new method needs to be applied. 
In the case of many-coupled chains, forming a 2D or 3D array, mean-field
approaches are, most likely, adequate. Note that this analysis can be
extended, in a straightforward way, to a many-chain system. This
crossover has been studied for a variety of interactions
in Ref. \cite{Tremblay}.

The most commonly studied perturbation has been the interchain hopping.
Then the required exponents are $\varsigma_{\uparrow}$ and
$\varsigma_{\downarrow}$ which are given in Eq. $(26)$. 
These operators are always relevant except as
$H \rightarrow H_c$. The fixed point which describes an isolated
Luttinger liquid is destabilized and interchain phase coherence develops. 
Note, however, that as in the presence of the magnetic field
$\varsigma_{\uparrow} \ne \varsigma_{\downarrow}$, 
the interchain hopping becomes spin dependent at low energies. 
This effect should influence the transport properties. It would be
interesting to check this fact experimentally.

So far, he have ignored the influence of the possible couplings among
themselves. To leading order in the strength of the couplings, this
is, at least, a second-order process, while the effect which leads to
(\ref{scaling}) is first order. It may happen, however, that new
couplings are generated, which were zero at the initial stages of the
renormalization process, i. e., in the bare Hamiltonian. When these new
couplings are more relevant than the ones initially present, they may
change drastically the physics at low energies. 

The coupling most likely to give rise to new interactions is the
interchain hopping. It involves two electronic operators, and,
to next order, it can generate four fermion operators, like 
interchain magnetic exchange, or Cooper pair hopping. In the following,
we analyze the process by which an exchange term is generated if,
initially, interchain hopping is present. We consider the
transverse-spin exchange
coupling, as this is the other most relevant operator in the presence of
electron-electron repulsion, besides the hopping. From the calculations 
presented in the preceeding sections, it may become more relevant than 
the hopping itself in the presence of a magnetic field.
The full scaling equation for this coupling, $J_{s \perp}$, is

\begin{equation}
\omega_c \frac{\partial {\tilde{J}}_{s \perp}}{\partial \omega_c}
= ( - 1 - 2 \varsigma_{s \perp}  ) {\tilde{J}}_{s \perp} -
\Gamma {\tilde{J}}_{\uparrow} {\tilde{J}}_{\downarrow} \, ,
\label{scalingx}
\end{equation}
where $\Gamma$ is a constant of order unity. The second term in the 
right-hand side stands for the fact that a simultaneous hopping of 
an up-spin electron in one direction and a down-spin electron in the 
opposite generates a spin-exchange process.

We can integrate Eqs. (\ref{scaling},\ref{scalingx}) with the initial
conditions: ${\tilde{J}}_{\uparrow} = {\tilde{J}}_{\downarrow} =
{\tilde{J}}_0$ and ${\tilde{J}}_{s \perp} = 0$. The scale-dependent 
coupling ${\tilde{J}}_{s \perp}$ depends on $l = \log ( \omega /
\omega_c )$ as

\begin{equation}
{\tilde{J}}_{s \perp} ( l ) \approx \frac{\Gamma {\tilde{J}}_{\uparrow}
{\tilde{J}}_{\downarrow} \left( e^{( 2 + 2 \varsigma_{\uparrow} +
2 \varsigma_{\downarrow} ) l } - e^{( 1 + 2 \varsigma_{s \perp} ) l} \right)}
{1 + 2 \varsigma_{\uparrow} + 2 \varsigma{\downarrow} - 2 \varsigma_{s
\perp}} \, .
\label{flowx}
\end{equation}

The low-energy behavior of the coupled system shows two different
regimes, depending on the value of the denominator in the right-hand 
side term in Eq. (\ref{flowx}):

i) If the denominator is of order unity, the interchain hopping diverges
first, at a scale given by Eq. (\ref{crossover}). Below this scale, a
system of coupled chains 
resembles a 2D or 3D anisotropic Fermi liquid, with a residual exchange
interaction. A mean-field analysis of this model leads to the
possibility of a SDW ground state.

ii) When the denominator is much less than one, the induced 
transverse-spin exchange 
grows faster than the hopping term. Then, before coherence between the
single-electron wavefunction at the different chains is established, the
exchange-term becomes of the order of the cutoff. The system probably
develops a spin gap, as in the previous case, but the transport
properties should be drastically different. 

Note, finally, that this analysis is valid for arbitrary strength of
the repulsion, $U$, and weak interchain couplings. It would
be interesting to connect this picture to the behavior obtained in the
opposite limit of strong-interchain hopping and weak-interchain
coupling, studied recently \cite{Fisher,Schulz2}.

\section{CONCLUDING REMARKS}

In this paper we have presented a comparative study of the
Hubbard chain instabilities. In contrast to the one-electron
1D quantum problem, the many-electron instabilities are not 
associated in general with logarithmic divergences in the
low-frequency response functions. Instead these
functions show Luttinger-liquid power laws which
are controlled by non-classical exponents. Moreover,
the singlet and triplet superconductivity instabilities
are removed by the effects of the electron correlations.

We have studied in detail the $U$, electronic density $n$, and
magnetic-field $H$ dependence of all the exponents which are 
associated with instabilities of the many-electron quantum 
problem. The interplay of the magnetic field and electron
correlations renders the transverse-spin SDW instability at 
momenta $\pm 2k_F$ dominant for large regions of parameter space. 
By removing the $SU(2)$ spin symmetry the magnetic field
also makes the exponents associated with the spin
and transverse-spin response functions different.

We went beyond CFT and characterized divergences which
occur at finite energy values in some correlation
functions in the cases of the half-filling
and fully-polarized ferromagnetic phases. We
have also studied the discontinuities which
occur in some power-law exponents which have
different values in both sides of the transitions
to or from the above phases.

Our exact results might be useful for the study
of the electronic instabilities in real quasi-one
dimensional materials. For instance, transverse SDW's
phases were often observed in real low-dimensional
materials \cite{Ishiguro,Gruner}. Following the discussion
of Sec. VI, elsewhere we will use the present instabilities of the 1D 
problem to construct a suitable effective Hamiltonian for the
description of a system of weakly coupled Hubbard chains. 

\nonum
\section{ACKNOWLEDGMENTS}

We are grateful to Peter Horsch for providing computer programs
which were useful for the production of our figures and for
stimulating discussions. We thank N. M. R. Peres and K. Maki for 
illuminating discussions. This research was supported in part by the 
Institute for Scientific Interchange Foundation under the EU Contract 
No. ERBCHRX - CT920020. PDS acknowledges partial support in the form 
of a PRAXIS XXI Fellowship.


$^{*}$ On leave of absence from Departamento de F\'{\i}sica, Instituto 
Superior T\'{e}cnico, Av. Rovisco Pais, P-1096 Lisboa Codex, Portugal.

\newpage

\begin{tabular}[b]{|l|l|l|l|l|l|l|}
\hline
& $U\rightarrow 0$ & \hspace{0.4cm}$H\rightarrow 0$ & 
$m\rightarrow 0,U\rightarrow\infty$ & \hspace{0.4cm}$m\rightarrow n$
& $H\rightarrow H_c, U\rightarrow\infty$ & \\
\hline
\hline
$\varsigma_{s\perp}$ & \hspace{0.4cm}$0$ & 
\hspace{0.4cm}$-1+{(\xi_0)^2\over 2}$ 
& \hspace{0.6cm}$-{1\over 2}$ & 
\hspace{0.3cm}$-1+(\eta_0)^2$ & \hspace{0.6cm}$-1$ & \\ 
\hline 
$\varsigma_{cs\downarrow}$ & \hspace{0.4cm}$0$ & 
\hspace{0.4cm}$-1+{(\xi_0)^2\over 2}$ 
& \hspace{0.6cm}$-{1\over 2}$ & \hspace{0.8cm}$0$ & \hspace{0.8cm}$0$ & \\ 
\hline 
$\varsigma_{cs\uparrow}$ & \hspace{0.4cm}$0$ & 
\hspace{0.4cm}$-1+{(\xi_0)^2\over 2}$ 
& \hspace{0.6cm}$-{1\over 2}$ & \hspace{0.3cm}$2[1-\eta_0]^2$ & 
\hspace{0.6cm}$+2$ & \\ 
\hline 
$\varsigma_{ss+}$ & \hspace{0.4cm}$0$ & \hspace{0.4cm}$2[{\xi_0\over 2}-{1\over 
{\xi_0}}]^2$ & \hspace{0.6cm}$+{1\over 2}$ & \hspace{0.3cm}$[1-\eta_0]^2$ 
& \hspace{0.6cm}$+1$ & \\ 
\hline 
$\varsigma_{ss-}$ & \hspace{0.4cm}$0$ & 
\hspace{0.4cm}$-1+{2\over (\xi_0)^2}$ & 
\hspace{0.6cm}$+1$ & $-1+4[1-{\eta_0\over 2}]^2$ & \hspace{0.6cm}$+3$ \\ 
\hline 
$\varsigma_{ts\downarrow}$ & \hspace{0.4cm}$0$ & 
\hspace{0.4cm}$-1+{2\over (\xi_0)^2}$ 
& \hspace{0.6cm}$+1$ & \hspace{0.3cm}$2[1-\eta_0]^2$ & 
\hspace{0.6cm}$+2$ & \\ 
\hline 
$\varsigma_{ts\uparrow}$ & \hspace{0.4cm}$0$ & 
\hspace{0.4cm}$-1+{2\over (\xi_0)^2}$ 
& \hspace{0.6cm}$+1$ & \hspace{0.8cm}$0$ & \hspace{0.8cm}$0$ & \\ 
\hline 
$\varsigma_{\uparrow}$ & \hspace{0.2cm}$-1$ & $-2 + 
{1\over 2}[{\xi_0\over 2}+{1\over {\xi_0}}]^2$ & 
\hspace{0.6cm}$-{7\over 8}$ & $-1+{1\over 2}[1-
\eta_0]^2$ & \hspace{0.6cm}$-{1\over 2}$ & \\ 
\hline 
$\varsigma_{\downarrow}$ & \hspace{0.2cm}$-1$ & $-2 + 
{1\over 2}[{\xi_0\over 2}+{1\over {\xi_0}}]^2$ & 
\hspace{0.6cm}$-{7\over 8}$ & $-1+{1\over 2}[1-\eta_0]^2$ 
& \hspace{0.6cm}$-{1\over 2}$ & \\
\hline 
\end{tabular}

TABLE -- Limiting values of the exponents whose
expressions are defined in Eqs. $(18)$, $(20)$, $(22)$, 
$(24)$, and $(26)$. Here the $m\rightarrow 0$ parameter 
$\xi_0=\xi_{cc}^1$ changes from $\xi_0=\sqrt{2}$ at $U=0$ to 
$\xi_0=1$ as $U\rightarrow\infty$ and the parameter $\eta_0$ is 
given in Eq. $(38)$. 

\newpage           

\figure{The exponents (a) $\varsigma_{s\perp}$, (b) 
$\varsigma_{cs\downarrow}$ and $\varsigma_{cs\uparrow}$, (c) 
$\varsigma_{\downarrow}$, and (d) $\varsigma_{\uparrow}$ defined 
by Eqs. $(18)$, $(20)$, and $(26)$ as functions of $U$ for $n=0.5$ 
and for values of the magnetic field $h=0.3$ (full line), $h=0.6$
(dashed line), and $h=0.9$ (dashed-dotted line).
\label{fig1}}

\figure{The exponents (a) $\varsigma_{ss+}$, (b) $\varsigma_{ss-}$,
(c) $\varsigma_{ts\downarrow}$, and (d) $\varsigma_{ts\uparrow}$
defined by Eqs. $(22)$ and $(24)$ as functions of $U$ for $n=0.5$ 
and for values of the magnetic field $h=0.3$ (full line), $h=0.6$
(dashed line), and $h=0.9$ (dashed-dotted line).
\label{fig2}}

\figure{The exponents (a) $\varsigma_{s\perp}$, (b) 
$\varsigma_{cs\downarrow}$ and $\varsigma_{cs\uparrow}$, (c) 
$\varsigma_{\downarrow}$, and (d) $\varsigma_{\uparrow}$ defined 
by Eqs. $(18)$, $(20)$, and $(26)$ as functions of the electronic density 
$n$ for $U=10$ and for values of the magnetic field $h=0.1$ 
(full line), $h=0.2$ (dashed line), and $h=0.3$ (dashed-dotted line).
\label{fig3}}

\figure{The exponents (a) $\varsigma_{ss+}$, (b) $\varsigma_{ss-}$,
(c) $\varsigma_{ts\downarrow}$, and (d) $\varsigma_{ts\uparrow}$
defined by Eqs. $(22)$ and $(24)$ as functions of the electronic 
density $n$ for $U=10$ and for values
of the magnetic field $h=0.1$ (full line), $h=0.2$
(dashed line), and $h=0.3$ (dashed-dotted line).
\label{fig4}}

\figure{The exponents (a) $\varsigma_{s\perp}$, (b) 
$\varsigma_{cs\downarrow}$ and $\varsigma_{cs\uparrow}$, (c) 
$\varsigma_{\downarrow}$, and (d) $\varsigma_{\uparrow}$ defined 
by Eqs. $(18)$, $(20)$, and $(26)$ as functions of the magnetic field $h$ 
for $n=0.5$ and various values of $U$.
\label{fig5}}

\figure{The exponents (a) $\varsigma_{ss+}$, (b) $\varsigma_{ss-}$,
(c) $\varsigma_{ts\downarrow}$, and (d) $\varsigma_{ts\uparrow}$
defined by Eqs. $(22)$ and $(24)$ as functions of the magnetic field $h$ 
for $n=0.5$ and various values of $U$.
\label{fig6}}

\figure{The exponents $\varsigma_{s\perp}$ and $\varsigma_{\uparrow}$
associated with the two dominant instabilities 
as functions of the magnetic field $h$ and for several values of the 
electronic density $n$ for (a) $U=1$, (b) $U=3$, (c) $U=5$, 
and (d) $U=20$. Comparison of the values
of both exponents shows that for intermediate and large
values of $U$ the transverse SDW becomes dominant for
magnetic fields $h$ larger than about $0.75$ to $0.8$.
\label{fig7}}

\figure{The critical line for $U_c^*(n)$ in the
$U-n$ plane for values of the magnetic field $h=0.8$ 
(full line) and $h=0.9$ (dashed line).
\label{fig8}}
\figure{The critical line for $h_c^*(n)$ in the $h-n$ plane
for values $U=10$ (full line) and $U=20$ (dashed line). 
\label{fig9}}
\end{document}